\documentclass[11pt,a4paper]{article}
\pdfoutput=1
\usepackage{jheppub}
\usepackage{rotating}
\usepackage{array}
\usepackage{amsmath,amssymb,amstext}
\usepackage[normalem]{ulem}
\usepackage{slashed}
\usepackage{booktabs}
\usepackage[pdftex,table,dvipsnames]{xcolor}
\usepackage{units}
\usepackage{multirow}
\usepackage{url}
\usepackage{color}
\usepackage{xspace}
\usepackage{comment}
\usepackage{enumitem}

\usepackage{multirow}
\preprint{P3H-22-043, TTK-22-16}

\title{Boosting mono-jet searches with model-agnostic machine learning}

\author{Thorben Finke,}
\author{Michael Kr\"amer,}
\author{Maximilian Lipp and}
\author{Alexander M\"uck}

\affiliation{Institute for Theoretical Particle Physics and Cosmology (TTK),\\ RWTH Aachen University, D-52056 Aachen, Germany}

\emailAdd{finke@physik.rwth-aachen.de}
\emailAdd{mkraemer@physik.rwth-aachen.de}
\emailAdd{maximilian.lipp@rwth-aachen.de}
\emailAdd{mueck@physik.rwth-aachen.de}

\abstract{We show how weakly supervised machine learning can improve the sensitivity of LHC mono-jet searches to new physics models with anomalous jet dynamics. The Classification Without Labels (CWoLa) method is used to extract all the information available from low-level detector information without any reference to specific new physics models. For the example of a strongly interacting dark matter model, we employ simulated data to show that the discovery potential of an existing generic search can be boosted considerably.}

\begin{document}

\maketitle

\section{Introduction}
\label{sec:Introduction}

The LHC is a designated discovery machine. It has delivered the discovery of a Standard Model-like Higgs particle~\cite{Aad:2012tfa,Chatrchyan:2012ufa} almost ten years ago and has also provided a wealth of precision measurements. However, new physics beyond the Standard Model (SM) has not been found in direct searches, yet. This is in contrast to our hope that there is a link between LHC physics and dark matter. The LHC experiments will increase the amount of collected data in the coming years, and even more so in the high luminosity phase of the LHC. This vast amount of data has to be scrutinized in all possible ways, in particular also expecting the unexpected.

Model-agnostic Machine Learning (ML) techniques have been shown to provide sensitivity to various new physics signatures. Completely unsupervised methods based on learning a representation of the data to find anomalies in a completely data-driven way have been introduced in Refs.~\cite{DeSimone:2018efk,Hajer:2018kqm,Heimel:2018mkt,Farina:2018fyg,Knapp:2020dde,Atkinson:2021nlt,Canelli:2021aps}. Guiding ML methods with simulations while keeping an open mind to what new physics might look like has been used in Refs.~\cite{Aguilar-Saavedra:2017rzt,DAgnolo:2018cun,DAgnolo:2019vbw,Amram:2020ykb,Aguilar-Saavedra:2020uhm}. It is the ultimate vision to have a model-agnostic ML algorithm that uncovers new physics even if it has never been considered by any human physicist. However, there is a long way to go. Being able to tag potential signal events, e.g.\ with autoencoders or any other anomaly search algorithm, does not mean that one can make a statistically meaningful discovery~\cite{Khosa:2022vxb} if there is no strategy to compare to the SM expectation. Only recently, first steps have been made to develop such a general, complete strategy~\cite{Mikuni:2021nwn}.

It is at the heart of any LHC new physics search to compare the background expectation with a measurement. If the background expectation can be measured in a control region (usually involving some theory assumptions), only data needs to be examined for potential discoveries.
ML methods have been proposed to increase the sensitivity of searches employing control regions~\cite{Dery:2017fap,Metodiev:2017vrx,Komiske:2018oaa,Collins:2018epr,Collins:2019jip,Nachman:2020lpy,Andreassen:2020nkr,Benkendorfer:2020gek,Dahbi:2020zjw,Lee:2019ssx,Hallin:2021wme,Raine:2022hht} and applied to experiment~\cite{Aad:2020cws}.
In a traditional resonance search or bump hunt, one simply counts events in side bands and signal regions and looks for deviations from a smooth distribution. In such a setup, model-agnostic ML is particularly promising since it can go far beyond counting events. ML can discover patterns in the signal region which are absent in the control regions, e.g.\ investigating jet substructure in the dijet invariant mass spectrum. To examine the potential of in this context, two community challenges have recently collected many interesting ideas~\cite{Aarrestad:2021oeb,Kasieczka:2021xcg}.

In this work we employ the Classification Without Labels (CWoLa) method \cite{Metodiev:2017vrx} which is designed to uncover any difference between a control and a signal region. A supervised classifier is trained to tag any event as belonging to the signal or the control region. If the control and the signal region have identically distributed features in the absence of new physics this is supposed to be an impossible task. The events being tagged as most signal like will be equally distributed between the control and the signal region. However, if there is a different signature in the signal region due to new physics, the supervised classifier will recognize it and tag it as most signal like. Moreover, the method can provide a statistically meaningful discovery. Labels for new physics events, which are unavailable in real data, are not needed. Only a control region is needed which contains less new physics events than the signal region.

CWoLa has been successfully used for bump hunts~\cite{Collins:2018epr,Benkendorfer:2020gek} and combined with density estimation to improve the sensitivity of the corresponding searches \cite{Nachman:2020lpy}. In the present work, we go beyond bump hunting. We show how to improve mono-jet searches at the LHC using CWoLa. The dominant background in a mono-jet search stems from invisibly decaying weak vector bosons. Hence, events with visibly decaying vector bosons naturally provide a suitable control region. A standard mono-jet search is a cut-and-count search, where features such as the structure of the observed jets do not play a role. Looking for any difference in the jet structure can boost the sensitivity to models with modified jet dynamics. This has been shown for a specific model using supervised ML for example in Ref.~\cite{Bernreuther:2020vhm}. Here, we show how the use of CWoLa improves the sensitivity to differences in jet structure in a model unspecific way. We demonstrate the general idea following the most recent ATLAS mono-jet search~\cite{ATLAS:2021kxv}. The discovery potential for new physics is illustrated employing a specific dark matter model with a strongly interacting dark sector~\cite{Bernreuther:2019pfb,Bernreuther:2020vhm}, highlighting the potential as well as the limitations of the method. In particular, we demonstrate that signal regions which are already limited by systematic uncertainties in a standard search, gain sensitivity using model-agnostic ML and will profit in particular from the high luminosity phase at the LHC.

This work is organised as follows: In Sec.~\ref{sec:Setup}, we introduce the ATLAS mono-jet search and describe the simulation of all the required data to be used in the following investigations, i.e.\ the SM backgrounds and the strongly interacting dark-matter model used as an example. Sec.~\ref{sec:CWoLamethod} summarizes the general CWoLa idea, before the specific CWoLa setup for the mono-jet search is described in Sec.~\ref{sec:CWoLahunting}. In Sec.\ref{sec:Proofofprinciple}, we highlight the discovery potential of the method under simplifying assumptions, before we scrutinize and reinforce our findings in Sec.~\ref{sec:Realitychecks}. We conclude in Sec.~\ref{sec:Conclusion} and refer to Appendices~\ref{sec:appendix_DGCNN} and \ref{sec:appendix_classifier} for more details on the setup and the results, respectively.

\section{Setup}
\label{sec:Setup}

The strategy to search for new physics as outlined in this paper is not tied to a specific mono-jet analysis, and it might also be promising for searches with a well-defined control region beyond a mono-jet analysis. However, for concreteness, we follow the ATLAS mono-jet search in Ref.~\cite{ATLAS:2021kxv}. In this search, events with $E_T^{\mathrm{miss}}> 200\, \text{GeV}$ and a leading anti-$k_T$ jet with jet radius 0.4, $p_T^{\mathrm{jet}}> 150\, \text{GeV}$ and $|\eta^{\mathrm{jet}}| < 2.4$ are analysed using an integrated luminosity of $139 \,  \mathrm{fb}^{-1}$ of LHC Run 2 in several inclusive and exclusive $E_T^{\mathrm{miss}}$ regions. 
At most three additional jets with $p_T^{\mathrm{jet}} > 30\, \text{GeV}$ and $|\eta^{\mathrm{jet}}| < 2.8$ are allowed in an event. Events with identified leptons are vetoed. We focus on the inclusive signal region IM1 with $E_T^{\mathrm{miss}}> 250\, \text{GeV}$ and $\Delta \phi(\mathbf{p}_T^\mathrm{jet}, \mathbf{p}_T^{\mathrm{miss}}) > 0.4$ for all jets. For this region more than $10^6$ events have been recorded. The choice of signal region IM1 is motivated by the fact that the CWoLa method benefits from large amounts of training data and will in particular improve searches which are systematics limited. In the high-luminosity phase of the LHC, our machine-learning approach will become more and more applicable also at higher $E_T^{\mathrm{miss}}$ thresholds. Other machine learning applications for improving the sensitivity of mono-jet searches can be found in Refs.~\cite{Dey:2019lyr,Khosa:2019kxd,Arganda:2021azw}.

\subsection{SM backgrounds}
\label{sec:Setup_SM}

The SM background in the signal region IM1 consists mainly of Z+jet production where the Z-boson decays invisibly into neutrinos (61\%), followed by W+jet production where the W-boson decays invisibly, i.e.\ it decays leptonically and the charged lepton is not identified (31\%). Smaller backgrounds are associated with top-quark production where one top quark decays leptonically (3.5\%) and diboson production where one boson decays invisibly (2\%). 

To reduce the systematic errors of the theory prediction, the background estimate of the ATLAS search~\cite{ATLAS:2021kxv} uses dedicated control samples where charged leptons from the Z-boson or W-boson decays are identified. The control region for IM1 is defined by the same cuts as the signal region, but the cut on $E_T^{\mathrm{miss}}$ is replaced by a cut on the recoil momentum $p_T^{\mathrm{recoil}}$, where $\mathbf{p}_T^{\mathrm{recoil}}=\mathbf{p}_{T}^l+\mathbf{p}_T^{\mathrm{miss}}$ and $\mathbf{p}_{T}^l$ is the transverse momentum of the identified leptons (one for W-boson decays, two for Z-boson decays). Whether an event belongs to the signal or the control region therefore only depends on the decay products of the vector boson. We make use of this control sample in our CWoLa approach as discussed in Sec.~\ref{sec:CWoLahunting}.

Since we do not have access to LHC data, we investigate our CWoLa based approach using Monte-Carlo simulations. All events are simulated at 13~TeV center-of-mass energy using MadGraph5~\cite{madgraph} for the hard process, Pythia8.2~\cite{Sjostrand:2014zea} as a parton shower, Delphes3~\cite{deFavereau:2013fsa} for a basic detector simulation using the ATLAS Card, and FastJet~\cite{Cacciari:2011ma} for jet-clustering, with the default settings for all tools. 

The CWoLa method is sensitive to differences in jet structure, comparing the jets in the control and the signal region. The leading QCD jets in the Z+jet and W+jet background processes are generated through initial-state radiation (ISR), so their structure is determined by QCD dynamics and independent of the decay channel and the decay dynamics of the vector boson. If the Z boson in the Z+jet process decays into neutrinos or charged leptons, the corresponding ISR QCD jet belongs to the signal or the control region, respectively. Also for W+jet production, the leading jets from ISR can populate the signal or the control region, depending on whether the charged lepton from the W~decay is identified or not. As the structure of the ISR jets is the same for all Z+jet and W+jet background processes, we simulate only Z+jet production followed by a Z decay into neutrinos and use the corresponding jets for both the signal and the control region. These events are the basis for the discussion in Sec.~\ref{sec:Proofofprinciple} where we show the applicability of the CWoLa method in a simplified setup.

As discussed above, the signal region also contains smaller admixtures from top-quark and diboson production. Fat jets (with a jet-radius $R = 0.8$, see Sec.~\ref{sec:CWoLahunting}) in those processes might have a multi-prong structure since they can contain more than one subjet emerging from the decay products of a top quark or a vector boson. Since the CWoLa method tags those jets as different if they are not present in the control region as well, they have to be treated with some care as discussed in Sec.~\ref{sec:Realitychecks}. To keep things simple, we only simulate $t\bar t$ production in the semi-leptonic channel, where the transverse momentum of the leptonically decaying W-boson has to fulfill $p_T^{\mathrm{recoil}}> 250\, \text{GeV}$. The leading jets from those $t\bar t$ events can populate the signal and the control region, depending on whether the charged lepton is identified or not. Concerning the diboson events, we simulate W($\to$ jets)Z$(\to\nu\bar\nu)$ events in the IM1 region. Again, we assume that a more realistic composition of the diboson events in the signal and control regions would not qualitatively change the jet characteristics of the sample we are interested in. Other backgrounds are strongly suppressed and are thus ignored in this work for the sake of simplicity.

\subsection{New physics example: Strongly interacting dark matter model}
\label{sec:newphysics}

The CWoLa strategy outlined in Sec.~\ref{sec:CWoLamethod} and Sec.~\ref{sec:CWoLahunting} is completely model agnostic. However, to show the discovery potential, we employ a specific new physics model. Since we are sensitive to modified jet dynamics, we use the strongly interacting dark matter model introduced in Ref.~\cite{Bernreuther:2019pfb}. How the modified jet dynamics of this model can be used to improve a mono-jet analysis in a supervised setup has been discussed in Ref.~\cite{Bernreuther:2020vhm}. Other machine learning applications to strongly interacting dark matter models can be found in Refs.~\cite{Park:2017rfb,Canelli:2021aps,Beauchesne:2021qrw}. For our choice of parameters, the model contains a heavy vector boson $Z'$ with mass $m_{Z'}=2~\mathrm{TeV}$ interacting with standard model quarks with a coupling $g_{\rm d}=0.1$. It couples the Standard Model to a sector of dark quarks $q_{\rm d}$ with a coupling $e_{\rm d}=0.4$. These quarks are also charged under a dark SU(3) gauge group with a confinement scale $\Lambda_{\rm d}=5\, \mathrm{GeV}$. Hence, being pair-produced in a $Z'$ decay they shower and hadronize to form dark pions $\pi_{\rm d}$ and dark rho mesons $\rho_{\rm d}$ with masses $m_{\pi_{\rm d}}= m_{\rho_{\rm d}}=5\, \mathrm{GeV}$. While the neutral rho mesons $\rho^0_{\rm d}$ mix with the $Z'$ and decay back to SM quarks promptly, the other mesons are stable dark-matter candidates and escape the detector. Hence, the jets in these models are usually called semi-visible jets. On average the invisible fraction of the jet energy amounts to $r_\mathrm{inv}=0.75$, leading to a specific modified jet structure. Furthermore, the dark jets differ from ordinary QCD jets due to the different running of the dark gauge coupling, the absence of heavy quarks in the shower and the presence of substructure from the decays of dark mesons. Because of the invisible fraction, the $Z'$ decays into dark quarks do not lead to a resonance in the dijet invariant mass. Moreover, a certain fraction of jets turns out to be completely invisible and populates the signal region of our mono-jet search. 

We use the UFO~\cite{Degrande:2011ua,Christensen:2008py} implementation of the model to simulate the pair-production of dark quarks starting with Madgraph and using the same tool-chain as discussed in Sec.~\ref{sec:Setup_SM}. The hidden-valley module of Pythia~\cite{Carloni:2010tw} is used to handle the dark showering, hadronisation and the decay of the $\rho^0_{\rm d}$ mesons into SM quarks. The subsequent showering and hadronisation of these quarks as well as the detector simulation are performed as for the SM backgrounds. Here, we simulate events in the IM1 region. The same model parameters and the same tool chain have been also used to produce the Aachen benchmark data set introduced in Ref.~\cite{Buss:2022lxw}.

\section{The CWoLa method}
\label{sec:CWoLamethod}

Classification Without Labels (CWoLa) is based on the typical setup in high energy physics experiments: One defines a signal region ($\rm SR$) and a control region ($\rm CR$). The signal region is chosen to optimize the fraction $f^{\rm SR}=N_{\mathrm{A}}^{\rm SR}/N_{\mathrm{B}}^{\rm SR}$ for a certain class of models, where $N_{\mathrm{A}}^{\rm SR}$ is the number of new physics or anomalous events/data instances and $N_{\mathrm{B}}^{\rm SR}$ is the number of background events/data instances. The control region should be free of anomalous events or should at least contain a smaller fraction $f^{\rm CR}=N_{\mathrm{A}}^{\rm CR}/N_{\mathrm{B}}^{\rm CR}$ of anomalous events, i.e.\ $f^{\rm CR}<f^{\rm SR}$. The expected number of background events $N_{\mathrm{B}}^{\rm SR}$ measured in the signal region is assumed to be known up to a certain relative error $\sigma$ which includes statistical as well as systematic uncertainties. The control region is used to minimize the systematic uncertainty using data driven methods. For $f^{\rm SR} > \sigma$ the measurement is sensitive, e.g. for $f^{\rm SR} > 5 \sigma$ one would expect a discovery.

There are features in each event that are used for the definition of the signal and the background region. CWoLa is based on the crucial assumption that there is an additional set of features such that background events from the signal and the control region are indistinguishable using only those features, i.e.\ the events in the signal and the control region are drawn from an identical probability distribution concerning this restricted feature space. These features will be called CWoLa features in the following.

In the CWoLa setup, a standard supervised binary classifier is trained on the CWoLa features which tags each event as belonging to the signal or the control region. These labels are available for real experimental data. No reference is made to labeling events as background or new physics events, which are the labels one would actually like to know, hence the name Classification Without Labels.

In the absence of new physics events and taking the above assumption for granted, the classifier has to fail because the task is impossible. The predicted labels cannot be better than random guessing. However, if there are anomalous events which are drawn from a different probability distribution and $f^{\rm CR}<f^{\rm SR}$, the classifier should assign a higher score for those anomalous events to belong to the signal region. Thus, for a given score threshold, the classifier will predominantly select anomalous events.

It is worth stressing the model-agnostic nature of this approach. Searching for a specific new physics model, a difference between the new physics and the background events in some feature will already be used to define the signal region to boost the sensitivity of a simple cut-and-count analysis. However, in this case one has sensitivity to this feature only. CWoLa instead is sensitive to all potential differences in the CWoLa features.

It has been shown that this setup leads to an optimal classifier~\cite{Metodiev:2017vrx}, i.e.\ it can have the same performance as a supervised classifier working with events labeled as being anomalous or belonging to the background. However, this finding does not imply that the method is guaranteed to work in practice.

CWoLa has been shown to improve the sensitivity of bump-hunt searches~\cite{Collins:2018epr}. In this context, the control region consists of the side bands of a specified signal region. For concreteness, consider a resonance search using dijet events. An interval in the dijet invariant mass is selected as signal region. It would be populated by new physics events if there was a resonance with a mass in the signal region decaying into a dijet final state. If the structure of new physics jets differs from QCD jets, the CWoLa method can boost the sensitivity of the search by only taking into account events with a classifier score beyond a certain threshold. However, the CWoLa features for the classifier have to be chosen with care since they might be correlated with the dijet mass which defines the signal region. Otherwise the assumption that background jets from the signal and the control region are indistinguishable using the CWoLa features is violated. The corresponding decorrelation of observables has been discussed in Ref.~\cite{Benkendorfer:2020gek} and methods to improve on those limitiations for bump hunts have also been suggested~\cite{Nachman:2020lpy}.

\section{CWoLa for anomalous mono-jets}
\label{sec:CWoLahunting}

To improve an existing or a future mono-jet search, we suggest to use the CWoLa setup in the following way: the signal region is defined as usual being mainly based on missing transverse energy. As discussed in Sec.~\ref{sec:Setup}, we use the IM1 region of the most recent ATLAS mono-jet search~\cite{ATLAS:2021kxv} as a concrete example. The control region is defined by events where neutrinos are replaced by charged leptons (see also Sec.~\ref{sec:Setup}). Data in the control regions has been recorded in past experimental analyses for most of those backgrounds to control systematics. Details can be also found in Ref.~\cite{ATLAS:2021kxv}.

As CWoLa features to the classifier, low level information of the leading fat jet in each event is used. To find the leading fat jet, the constituents of the events in the IM1 region are reclustered into anti-$k_T$ jets with a jet radius $R = 0.8$. In our simulation based studies, we use the 40 jet constituents with largest transverse momentum in the constituents branch of the Delphes output (see Sec.~\ref{sec:Setup} for the simulation details). The assumption that the CWoLa features are uncorrelated with respect to the definition of signal and background regions, i.e.\ the leptonic decays of weak bosons, is physically sound: the evolution of the jets from initial-state radiation is driven by QCD dynamics and not influenced by the decay properties of a leptonically decaying weak boson which recoils against the jet. We have verified the independence of the jet structure with respect to the weak-boson decays in our simulation and simplify the simulation accordingly as detailed in Sec.~\ref{sec:Setup}. Replacing simulated by experimentally recorded data should be straightforward.

As the binary classifier, we use a Dynamic Graph Convolutional Neural Network (DGCNN)~\cite{DGCNN} which is based on the ParticleNet architecture~\cite{Qu:2019gqs} and has proven to be an extremely powerful jet tagger~\cite{Kasieczka_top}. Architecture, preprocessing, and training are discussed in Appendix~\ref{sec:appendix_DGCNN}. Note that instead any powerful supervised classification algorithm could be used, e.g.\ the recently proposed graph-based LundNet~\cite{Dreyer:2020brq} or LorentzNet~\cite{Gong:2022lye}.

The score $s \in [0,1]$ of the trained classifier for each event can be interpreted as the probability of the event to belong to the signal region (see Appendix~\ref{sec:appendix_DGCNN} for details). For few anomalous events, the output distribution for background events is expected to peak close to $s=0.5$. As discussed in Sec.~\ref{sec:CWoLamethod}, anomalous events are expected to be tagged as belonging to the signal region.

Given the classifier score $s$ for each event, we choose a threshold $t$. For $s>t$, the event is selected as being potentially anomalous. The threshold $t$ is chosen such that one permille of the events in the control region is selected, i.e.\ $n^{\rm CR}=\epsilon^{\rm CR} N^{\rm CR}$ with $\epsilon^{\rm CR}=0.001$. We will comment on this choice below. Here we always assume that there is only background in the control region. Hence, we work with a background rejection $1/\epsilon^{\rm CR}_{\rm B}=1/\epsilon^{\rm CR}=1000$. For a given $1/\epsilon^{\rm CR}_{\rm B}$, corresponding to a specific threshold $t$, the classifier selects a number of anomalous events $n_{\mathrm{A}}^{\rm SR}=\epsilon^{\rm SR}_{\rm S}  N_{\mathrm{A}}^{\rm SR}$ in the signal region, where $\epsilon^{\rm SR}_{\rm S}$ is the signal efficiency and depends on $t$. In our weakly supervised setup, $\epsilon^{\rm SR}_{\rm S}$ is unknown. In analogy, the classifier selects a certain number of background events from the signal region $n_{\mathrm{B}}^{\rm SR}=\epsilon^{\rm SR}_{\rm B}  N_{\mathrm{B}}^{\rm SR}$. If the CWoLa assumptions hold, we have $\epsilon^{\rm SR}_{\rm B}=\epsilon^{\rm CR}_{\rm B}=0.001$.

If there are no anomalies in the signal region, the fraction of selected events from the signal region will be identical to that of the control region, i.e.\ the expected number of selected events is $n^{\rm SR}_\mathrm{exp}=\epsilon^{\rm CR}_{\rm B}  N^{\rm SR}$. This is the Null hypothesis. If $n^{\rm SR}_\mathrm{exp}$ and $n^{\rm SR}$ differ only as expected from statistical fluctuations, no indication of new physics is observed. If there are anomalous events in the signal region and the classifier is successful in identifying them with some $\epsilon^{\rm SR}_{\rm S}>\epsilon^{\rm SR}_{\rm B}$, the fraction of selected events from the signal region will be larger. If it exceeds the expectation for statistical fluctuations the Null hypothesis can be excluded in the usual way. Due to the model agnostic nature of the method, there are no exclusion limits to be derived for any models. The CWoLa method is a discovery tool.

The choice for the background rejection $1/\epsilon^{\rm CR}_{\rm B}$ is to a certain extent arbitrary. Our choice $1/\epsilon^{\rm CR}_{\rm B}=1000$ is driven by the following considerations: Although $\epsilon^{\rm SR}_{\rm S}$ is unknown, the signal-to-background ratio $n_{\mathrm{A}}^{\rm SR}/n_{\mathrm{B}}^{\rm SR}$ is usually a monotonically growing function of the background rejection $1/\epsilon^{\rm CR}_{\rm B}$ which favours to choose  $1/\epsilon^{\rm CR}_{\rm B}$ large. In particular, the signal-to-background ratio for a discovery should not be too small such that one is not too sensitive to small unknown systematics concerning the validity of the CWoLa assumptions. On the other hand, for increasing $1/\epsilon^{\rm CR}_{\rm B}$, a classifier is often not performant enough to not only improve the signal-to-noise ratio but also the significance $n_{\mathrm{A}}^{\rm SR}/\sqrt{n_{\mathrm{exp}}^{\rm SR}}$ of an observed excess. Hence, the minimal acceptable signal-to-background ratio for a discovery is a good guide for $1/\epsilon^{\rm CR}_{\rm B}$.

As a default, we use one million events in both the signal and the control region which is close to the actual number of observed events in the IM1 signal region of the most recent ATLAS search~\cite{ATLAS:2021kxv}. For $\epsilon^{\rm CR}_{\rm B}=0.001$, this corresponds to $n^{\rm CR}=1000$ selected events. Although $n^{\rm CR}$ is fixed by choosing the threshold $t$ for our data set, $n^{\rm CR}$ for the same $t$ is distributed with the usual statistical uncertainty $\sqrt{n^{\rm CR}}$ for independent data sets. Therefore the corresponding relative statistical uncertainty due to the limited size of the the control region for $n^{\rm SR}$ is roughly $\sigma^{\rm CR}=3\%$. Having a signal sample of the same size ($n^{\rm SR}_\mathrm{exp} = n^{\rm CR}$) and adding the corresponding relative statistical uncertainty $\sigma^{\rm SR}$ in quadrature, a $5 \sigma$ discovery needs at least $5 \sqrt{(\sigma^{\rm CR})^2 + (\sigma^{\rm SR})^2}\, n^{\rm SR}_\mathrm{exp} = 5 \sqrt{2}\, \sigma^{\rm CR} \,n^{\rm SR}_\mathrm{exp} = 5 \sqrt{2 \, n^{\rm SR}_\mathrm{exp}} \sim 224$ additional events.

In principle, one could also scan the background rejection. However, here we do not follow this idea in order to avoid discussions about the look-elsewhere effect. The chosen value has not been tuned to the success of finding our example signal.
Moreover, as we will see in the following sections, a discovery using our CWoLa setup will most likely be an iterative process which is driven not so much by statistical considerations but by iterated efforts to understand the quality of the control region.

In this setup we are sensitive to modified jet structures which occur in events with large missing transverse energy. A physically well motivated example for such a model is discussed in Sec.~\ref{sec:newphysics} and used in the following sections to demonstrate the sensitivity of the method.

\section{Proof of principle}
\label{sec:Proofofprinciple}

In this section, we take the CWoLa assumptions for granted. The simulated background events in the signal and the control region are indeed drawn from the same probability distribution since we use the same simulation setup to generate them. We only use Z+jet events which are simulated as discussed in Sec.~\ref{sec:Setup_SM}. Therefore, we investigate the performance of the CWoLa setup under ideal circumstances. The problems which might arise for more complicated samples of background jets or in defining a suitable control sample are discussed in Sec.~\ref{sec:Realitychecks}.

As a default, we use $N^{\rm CR}=N^{\rm SR}=10^6$ events in the control region and in the signal region. We use a fraction $f^{\rm CR}=0$ of anomalous events in the control region, i.e.\ we have $N^{\rm CR}_\mathrm{B}=N^{\rm CR}$ and only Z+jet events. In the signal region, we have $N^{\rm SR}_\mathrm{B}=(1-f^{\rm SR}) N^{\rm SR}$ Z+jet events and $N^{\rm SR}_\mathrm{A}=f^{\rm SR} N^{\rm SR}$ new physics events, simulated as detailed in Sec.~\ref{sec:newphysics}. We take the number of new physics events $N^{\rm SR}_\mathrm{A}$ as a free parameter. Note that not all new physics events have anomalous leading fat jets since also initial-state radiation QCD jets can be leading. This is making the task to identify new physics even harder.

Using $\epsilon^{\rm CR}_{\rm B}=0.001$, as discussed in Sec.~\ref{sec:CWoLahunting}, the number of events $n^{\rm SR}=n_{\mathrm{A}}^{\rm SR}+n_{\mathrm{B}}^{\rm SR}$ selected by the classifier in the signal region is shown in Tab.~\ref{tab:result1} for several signal fractions $f^{\rm SR}$. To be less vulnerable to fluctuations in the training process we train five classifiers on the same data and average their scores. Most importantly, the Null test works fine. If there are no anomalous events in the data ($f^{\rm SR}=0$), the selected number is, within statistical fluctuations, in agreement with the expected value $n^{\rm SR}_\mathrm{exp}=1000$. CWoLa does not provide any false indication for new physics. Hence, overfitting is no issue. Moreover, a signal rate $f^{\rm SR}=1\%$, which is still consistent with constraints from the latest ATLAS mono-jet search, leads to $n^{\rm SR}=1666$. Without statistical doubts, such a finding would indicate that there is something to be understood about the data. Ideally, a thorough investigation of the selected jets will uncover the unexpected jet structure and hint towards a suitable new physics model.

\begin{table}
  \begin{center}
    \begin{tabular}{c||c|c||c|c||c}
      $f^{\rm SR}$ & $n^{\rm SR}_\mathrm{exp}$ & $n^{\rm SR}$ & $n^{\rm SR}_\mathrm{A}$ & $n^{\rm SR}_\mathrm{B}$&$(n^{\rm SR}-n^{\rm SR}_\mathrm{exp})/\sqrt{2\, n^{\rm SR}_\mathrm{exp}}$\\
      \hline
      0\%     & 1000  & 1048  & 0     & 1048  & 1.07  \\
      0.2\%   & 1000  & 1065  & 47    & 1018  & 1.45  \\
      0.4\%   & 1000  & 1107  & 100   & 1007  & 2.39  \\
      0.5\%   & 1000  & 1175  & 184   & 991   & 3.91  \\
      0.6\%   & 1000  & 1306  & 247   & 1059  & 6.84  \\
      0.7\%   & 1000  & 1389  & 367   & 1022  & 8.70  \\
      0.8\%   & 1000  & 1500  & 419   & 1081  & 11.18 \\
      1\%     & 1000  & 1666  & 625   & 1041  & 14.89 \\
      2\%     & 1000  & 2357  & 1392  & 965   & 30.34 \\
      4\%     & 1000  & 4182  & 3269  & 913   & 71.15 \\
    \end{tabular}
  \end{center}
  \caption{\label{tab:result1} Number of events $n^{\rm SR}$ selected  from the signal region for $N^{\rm CR}=N^{\rm SR}=10^6$ and several signal fractions $f^{\rm SR}$. We have used the mean score of five classifiers trained on the same data. We also show the number of events expected to be selected in the absence of a new physics signal, $n^{\rm SR}_\mathrm{exp}$, and the number of selected anomalous and background events ($n^{\rm SR}_\mathrm{A}$ and $n^{\rm SR}_\mathrm{B}$). The latter two numbers are not known for real data. The last column shows an estimate of the statistical significance of a possible discovery.}
\end{table}

Around a signal fraction $f^{\rm SR}=0.6\%$, the statistical significance rapidly drops below $5\sigma$. Even under the ideal circumstances which are assumed in this chapter, the classifier is then not able to identify the new physics events as anomalous. Note that only $N^{\rm SR}_\mathrm{A}=6000$ anomalous events are in the training sample with $N^{\rm SR}+N^{\rm CR}=2\cdot 10^6$ events. It is extremely challenging for a classifier to efficiently learn the anomalous structures under these circumstances. In a supervised setup, the data instances would be weighted according to their abundance to help the classifier. In our weakly supervised setup, however, this is not possible.

Moreoever, our studies show that the absolute number of anomalous events is an essential parameter. If the number of background events is increased for a fixed number of anomalous events in the signal region, the performance of the CWoLa method is relatively stable, although the signal fraction is decreasing. In Tab.~\ref{tab:result2}, we show the tagged number of events from the signal region for several $N^{\rm SR}$, with the number of anomalous events fixed at $N^{\rm SR}_\mathrm{A}=10$k. Moreover, Tab.~\ref{tab:result3} shows the improving performance for a fixed signal fraction $f^{\rm SR} = 0.01$ if the collected data increase. This is good news for the high-luminosity phase of the LHC. Whether improved training strategies or more powerful classifiers can improve the overall performance is left for future research.

\begin{table}
  \begin{center}
    \begin{tabular}{c||c|c||c|c||c}
      $N^{\rm SR}$ & $n^{\rm SR}_\mathrm{exp}$ & $n^{\rm SR}$ & $n^{\rm SR}_\mathrm{A}$ & $n^{\rm SR}_\mathrm{B}$ &$(n^{\rm SR}-n^{\rm SR}_\mathrm{epx})/\sqrt{2\, n^{\rm SR}_\mathrm{epx}}$\\
      \hline
      $250$k  & 250   & 916   & 686   & 230   & 29.78 \\
      $500$k  & 500   & 1113  & 660   & 453   & 19.38 \\
      $1000$k & 1000  & 1666  & 625   & 1041  & 14.89 \\
    \end{tabular}
  \end{center}
  \caption{\label{tab:result2} Same as Tab.~\ref{tab:result1} but for a fixed number of anomalous events $N^{\rm SR}_\mathrm{A}=10$k in a signal region with $N^{\rm SR}$ events.}
\end{table}

\begin{table}
  \begin{center}
    \begin{tabular}{c||c|c||c|c||c}
      $N^{\rm SR}$ & $n^{\rm SR}_\mathrm{exp}$ & $n^{\rm SR}$ & $n^{\rm SR}_\mathrm{A}$ & $n^{\rm SR}_\mathrm{B}$ &$(n^{\rm SR}-n^{\rm SR}_\mathrm{epx})/\sqrt{2\, n^{\rm SR}_\mathrm{epx}}$\\
      \hline
      $250$k  & 250   & 362   & 126   & 236   & 5.01  \\
      $500$k  & 500   & 680   & 258   & 422   & 5.69  \\
      $1000$k & 1000  & 1666  & 625   & 1041  & 14.89 \\
    \end{tabular}
  \end{center}
  \caption{\label{tab:result3} Same as Tab.~\ref{tab:result1} but for a fixed fraction $f^{\rm SR}=0.01$ of anomalous events in a signal region with $N^{\rm SR}$ events.}
\end{table}

We have also investigated the effect of the control region being smaller than the signal region; in the analysis we have considered as an example, the smaller branching ratio of the Z boson 
into charged leptons leads to a smaller number of events in the control sample. Weighting the events accordingly, we do not observe a significant loss of discovery power beyond the increased statistical error.

\section{Reality checks}
\label{sec:Realitychecks}

In Sec.~\ref{sec:Proofofprinciple}, we have shown the sensitivity of the CWoLa method under idealized conditions. In this section, we study how a non-trivial and more realistic composition of the signal and control samples impacts the performance. Here, we assume that the background events in the signal region contain $r^{\rm SR}_{t\bar t}=3.5\%$ $t\bar t$ events and $r^{\rm SR}_{VV}=2\%$  diboson events (see Sec.~\ref{sec:Setup_SM}). In particular, we investigate how well this composition has to be understood and reflected in the control region.

\begin{table}
  \begin{center}
    \begin{tabular}{c|c||c|c|c||c|c|c||c}
      $r^{\rm CR}_{t\bar t}$ & $r^{\rm CR}_{VV}$ & $n^{\rm CR}_{\mathrm{Z+jet}}$ & $n^{\rm CR}_{t\bar t}$ & $n^{\rm CR}_{VV}$ &  $n^{\rm SR}_{\mathrm{Z+jet}}$ & $n^{\rm SR}_{t\bar t}$ & $n^{\rm SR}_{VV}$ & $n^{\rm SR}$ \\
      \hline
      0\%     & 0\%   & 1000  & 0     & 0     & 1053  & 1127  & 2043  & 4223  \\
      1.75\%  & 1.0\% & 541   & 247   & 212   & 581   & 557   & 420   & 1558  \\
      2.80\%  & 1.6\% & 980   & 3     & 17    & 1065  & 6     & 17    & 1088  \\
      3.15\%  & 1.8\% & 957   & 24    & 19    & 962   & 26    & 32    & 1020  \\ 
      \hline
      3.50\%  & 2.0\% & 823   & 160   & 17    & 793   & 175   & 21    & 989   \\ 
      \hline
      5.00\%  & 3.0\% & 966   & 18    & 16    & 960   & 22    & 3     & 985   \\
      3.50\%  & 1.6\% & 903   & 89    & 8     & 821   & 93    & 26    & 940   \\
      2.80\%  & 2.0\% & 983   & 5     & 12    & 996   & 7     & 12    & 1015  \\
    \end{tabular}
  \end{center}
  \caption{\label{tab:result4} Number of events selected by the classifier from the different background classes in the signal and control region for different compositions of the control region. There are no anomalous events ($f^{\rm SR}=0$) and we have $r^{\rm SR}_{t\bar t}=3.5\%$ and $r^{\rm SR}_{VV}=2\%$.}
\end{table}

In the absence of any new physics events ($f^{\rm CR}=f^{\rm SR}=0$), we show the results of the CWoLa tagging for several values of $r^{\rm CR}_{t\bar t}$ and $r^{\rm CR}_{VV}$ in Tab.~\ref{tab:result4}. For $r^{\rm CR}_{t\bar t}=r^{\rm CR}_{VV}=0$, i.e.\ the additional backgrounds are absent from the control region, the CWoLa method correctly identifies the different jet structures from the top-quark and weak-boson decays in the signal region. However, this is of course not a sign for new physics but a consequence of our poor modeling of the control region. Although this is a naive example, it highlights the challenge of the CWoLa approach: tagging more signal-region events than expected can always either be due to new physics or due to a mismodeling of the control region. For real data, the leading jets for the control region have to be measured from different event topologies and then combined to form a proper control region that matches the expected rates of the different backgrounds in the signal region, where input from theory and Monte Carlo is required. Therefore, it is a relevant question to which extent the control region needs to be understood. The results in Tab.~\ref{tab:result4} show, that the understanding and the inclusion of the backgrounds are crucial at the percent level, as it is already the case for the standard searches. However, variations of backgrounds between the control and signal region at the level of a few permille (i.e.\ a relative understanding at the level of 10\%) are tolerable. Hence, one does not need to be perfect. In particular, the CWoLa setup is not too sensitive to overestimating small background components.

Moreover, in a certain sense, we propose a self-correcting setup. Tagging more events from the signal region than expected in real data would first of all prompt more efforts to better understand the control region. The tagged events might also help in understanding which backgrounds are mismodeled. Only after all those studies one would want to pursue an interpretation of the selected signal events in terms of new physics, also aided by the investigation of the tagged jet's structure. 

In Tab.~\ref{tab:result5} we again study if the exemplary semivisible jets can be tagged, using a more realistic background sample in the signal as well as in the control region. We see that excesses in $n^{\rm SR}$ over the expected number of observations are dominated by these semivisible jets, even for slightly mismodeled control regions. Completely neglecting additional backgrounds in the control region, however, results in an excess dominated by these backgrounds and only minor enhancement of the fraction of signal jets.

In Appendix~\ref{sec:appendix_classifier}, we further discuss the classifier score using the Monte Carlo labels of the events which are not available for real data.

\begin{table}
  \begin{center}
    \begin{tabular}{c|c||c|c|c||c|c|c|c||c}
      $r^{\rm CR}_{t\bar t}$ & $r^{\rm CR}_{VV}$ & $n^{\rm CR}_{\mathrm{Z+jet}}$ & $n^{\rm CR}_{t\bar t}$ & $n^{\rm CR}_{VV}$ &  $n^{\rm SR}_{\mathrm{Z+jet}}$ & $n^{\rm SR}_{t\bar t}$ & $n^{\rm SR}_{VV}$ & $n^{\rm SR}_{A}$ & $n^{\rm SR}$ \\
      \hline
      0\%     & 0\%   & 1000  & 0     & 0     & 1089  & 1245  & 1826  & 223   & 4383  \\
      2.80\%  & 1.6\% & 876   & 108   & 16    & 838   & 134   & 37    & 456   & 1465  \\
      \hline
      3.50\%  & 2.0\% & 963   & 23    & 14    & 996   & 25    & 32    & 633   & 1686  \\
      \hline
      5.00\%  & 3.0\% & 971   & 12    & 17    & 1034  & 8     & 10    & 575   & 1627  \\
    \end{tabular}
  \end{center}
  \caption{\label{tab:result5} Number of events selected by the classifier from the different classes in the signal and control region for different compositions of the control region. There is a fixed fraction of anomalous events ($f^{\rm SR}=0.01$) and we have $r^{\rm SR}_{t\bar t}=3.5\%$ and $r^{\rm SR}_{VV}=2\%$.}
\end{table}

\section{Conclusion}
\label{sec:Conclusion}

We have demonstrated that the Classification Without Labels (CWoLa) method~\cite{Metodiev:2017vrx} is a powerful tool to boost the LHC discovery potential for new physics models with anomalous jet dynamics and a mono-jet signature. There are no model-specific assumptions, and the proposed setup can be implemented directly using collected data. The CWoLa method relies on a background-dominated control sample, which is in general available for LHC searches.

Using Monte Carlo simulation and a well-motivated specific new physics model with a strongly interacting hidden sector~\cite{Bernreuther:2019pfb}, we show that less than 1\% of new physics events in inclusive signal regions would be sufficient to discover signs of new physics. In contrast, the corresponding traditional search~\cite{ATLAS:2021kxv} is not even sensitive enough to exclude the model at 95\% confidence level (the systematics dominated error on the SM prediction for signal region IM1 is 1.2\%). We consider the hidden sector model to be a rather challenging test case since the modified jet dynamics of the model is difficult to recognize by unsupervised methods~\cite{Buss:2022lxw} and by dedicated supervised taggers~\cite{Bernreuther:2020vhm}.

The CWoLa setup, as a weakly supervised method, is not as sensitive as a dedicated model-specific and Monte Carlo-based approach using supervised methods. Moreover, as a discovery tool, it will not provide exclusion limits for specific model parameters. On the other hand, we emphasise again that the CWoLa method is model agnostic and can be applied directly to data. We have also demonstrated that the CWoLa method provides a useful data-driven tool to improve the understanding of the signal and control regions.

We have shown that the CWoLa discovery potential is increasing with more data, while traditional searches might already be limited by systematic uncertainties. The method should therefore continue to gain importance with, in particular, the large amounts of data expected in the high-luminosity phase of the LHC. Given that the CWoLa method is model-agnostic and data-driven and that it only requires a background-dominated control sample, we consider CWoLa-assisted searches for new physics and reanalyses of data already collected by the LHC experimental collaborations to be very promising.

\section*{Acknowledgements}

We would like to thank Elias Bernreuther for discussions and comments on the manuscript. TF is supported by the Deutsche Forschungsgemeinschaft (DFG, German Research Foundation) under grant 400140256 - GRK 2497: The physics of the heaviest particles at the Large Hadron Collider. The research of MK and AM is supported by the Deutsche Forschungsgemeinschaft (DFG, German Research Foundation) under grant 396021762 - TRR 257: Particle Physics Phenomenology after the Higgs Discovery. The authors gratefully acknowledge the computing time granted by the NHR4CES Resource Allocation Board and provided on the supercomputer CLAIX at RWTH Aachen University as part of the NHR4CES infrastructure. The calculations for this research were conducted with computing resources under the project rwth0934. 

\begin{appendix}

\section{The DGCNN classifier}
\label{sec:appendix_DGCNN}

As input for our Dynamic Graph Convolutional Neural Network (DGCNN) we use the 40 leading $p_T$ constituents of each jet, zero padded when needed. From the 4-momentum of each jet constituent we construct the set of seven input features $\{\Delta \eta$, $\Delta \phi$, $\log(p_T)$, $\log(p_T / p_{T}^{\text{jet}})$, $\log(E)$, $\log(E / E^\text{jet})$, $\Delta R\}$ with $\Delta \eta=\eta-\eta^\text{jet}$, $\Delta \phi=\phi-\phi^\text{jet}$ and $\Delta R=\sqrt{\Delta \eta ^2 + \Delta \phi ^2}$; $\eta$, $\phi$, $p_T$ and $E$ refer to the rapidity, the azimuthal angle, the transverse momentum (in GeV), and the energy (in GeV) of the constituent, respectively. The quantities with superscript ``jet" refer to the respective characteristics of the jet.

The DGCNN constructs a k-nearest-neighbors (knn) graph with these particles as nodes. We use $k=16$. The initial graph is constructed using the Euclidean distances of the particles in $\Delta \eta$ and $\Delta \phi$. The network's architecture is almost identical to the one used in Ref.~\cite{Bernreuther:2020vhm} for supervised classification on similar data sets, i.e.\ we use three EdgeConv blocks with three convolutions each. The number of features is increased successively from the seven input features to 64, 128 and finally 256. The graph is dynamically updated as knn graph after the first and second block using the Euclidean distances between nodes/particles based on all features. The outputs of each block are concatenated with the input resulting in 455 features per particle. After global average pooling the 455 features are fed into a fully connected network with 256, 128 and 2 nodes. We regularize the fully connected network using dropout with a fraction of 0.1 after the first two layers. We apply softmax activation to the last layer, allowing for a probability interpretation of the output. The two outputs then correspond to the probability of the input to belong to the control region or the signal region. We use the probability for the signal region as score $s$ (see Sec.~\ref{sec:CWoLahunting}). The only difference to Ref.~\cite{Bernreuther:2020vhm} is that we use Leaky ReLU with $\alpha = 0.1$ instead of ReLu as activation function, since we observed that it leads to more stable results.

We implement the network using TensorFlow 2.6.0~\cite{abadi2016tensorflow} and the build-in version of Keras~\cite{chollet2015keras}. We use the Adam optimizer~\cite{Kingma:2014vow} with its default settings to minimize the categorical cross entropy. During training we reduce the learning rate by a factor of 0.1 when the training loss does not improve for 8 epochs. If the loss still does not improve for an additional four epochs, we stop the training. We set the maximum number of training epochs to 75, which is hardly ever reached. Note that we evaluate the network on the same data that we use to train. This corresponds to the procedure one would use on experimental data. The results in Sec.~\ref{sec:Proofofprinciple} show that overfitting is not a problem for such a large dataset, as we see no excess in selected events in the signal region in the absence of signal (see Tab.~\ref{tab:result1}).

\section{Classifier output}
\label{sec:appendix_classifier}

In this appendix we show and analyze the classifier score for the training sample. Here, we make use of the Monte Carlo labels for anomalous/new physics events and background events of the different types. Hence, this information is not available for data collected at the LHC. However, it is nevertheless instructive to investigate it.

\begin{figure}[p]
  \centering
  \includegraphics[width=0.49\textwidth]{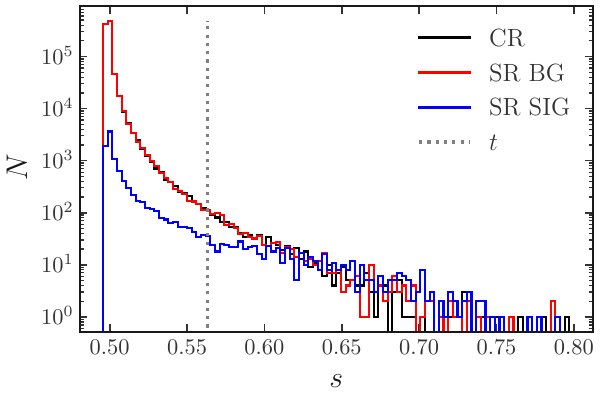}
  \includegraphics[width=0.49\textwidth]{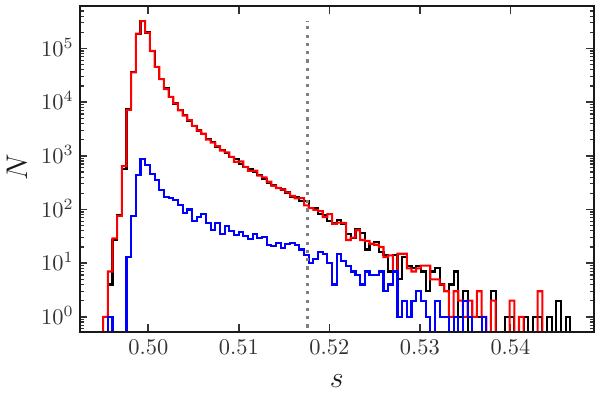}
  \caption{The absolute number of events in the signal region (Z+jet events in red, semivisible jets in blue) and the control region (only Z+jet events in black) is shown as a function of the classifier score $s\in[0,1]$ for $f^{\rm SR}=1\%$ (left) and $f^{\rm SR}=0.5\%$ (right). Note the different scale on the $x$-axis in the two plots. We also show the threshold value $t$ for $\epsilon^{\rm CR}=0.001$ as a vertical line.}
  \label{fig:app1}
\end{figure}

In Fig.~\ref{fig:app1}, we show the signal score $s$ for the simplified setup used in Sec.~\ref{sec:Proofofprinciple}. The scores peak at $s=0.5$, as this minimizes the loss function if all jets in the control and the signal region are drawn from identical probability distributions. For a well-defined control region there are no jets which are present only in the control region but not in the signal region, such that scores well below $s=0.5$ would be a sign of overfitting and should not be observed. In the signal region, we distinguish jets from Z+jet background events and semivisible jets from new physics events. For a signal fraction $f^{\rm SR}=1\%$ (and $f^{\rm CR}=0$), there are enough semivisible jets such that the training is successful in at least identifying the most anomalous jets. Most jets from new physics events are not identified, but this is not a problem for the CWoLa method to work. For $f^{\rm SR}=0.5\%$, those anomalous jets are still there, but the training procedure is not able to distinguish them in an efficient way. In particular the range of classifier scores is significantly reduced to $s<0.55$ ($s<0.8$ for $f^{\rm SR}=1\%$).

\begin{figure}[p]
  \centering
  \includegraphics[width=0.49\textwidth]{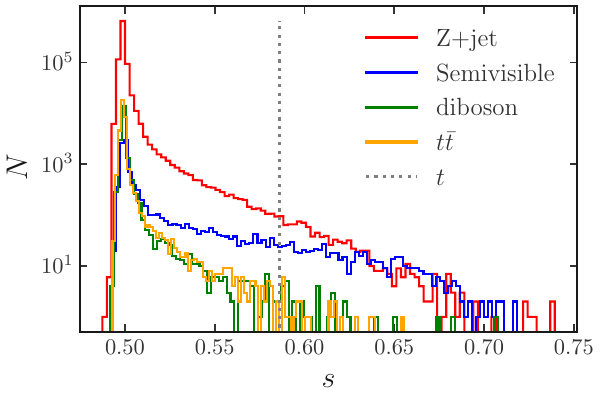}
  \includegraphics[width=0.49\textwidth]{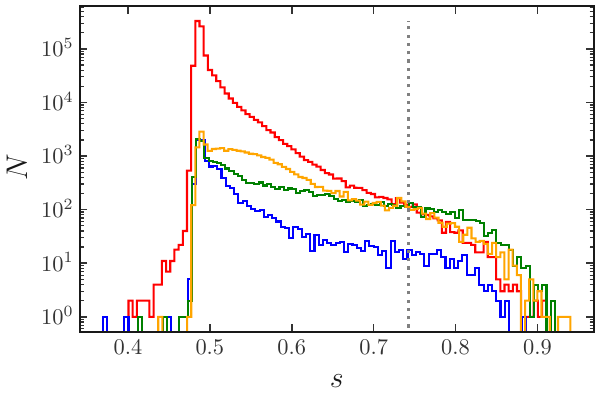}
  \caption{The absolute number of events in the signal region (Z+jet events in red, semivisible jets in blue, $t\bar t$ events in orange, diboson events in green) is shown as a function of the classifier score $s\in[0,1]$. We have $r^{\rm SR}_{t\bar{t}} = 3.5\%$ and $r^{\rm SR}_{VV} = 2.0\%$ and use $r^{\rm CR}_{t\bar{t}} = r^{\rm SR}_{t\bar{t}}$ and $r^{\rm CR}_{VV} = r^{\rm SR}_{VV}$ (left) or $r^{\rm CR}_{t\bar{t}} = r^{\rm CR}_{VV} = 0$ (right). We also show the threshold value $t$ for $\epsilon^{\rm CR}=0.001$ as a vertical line.}
  \label{fig:app2}
\end{figure}

\begin{figure}[p]
  \centering
  \includegraphics[width=0.49\textwidth]{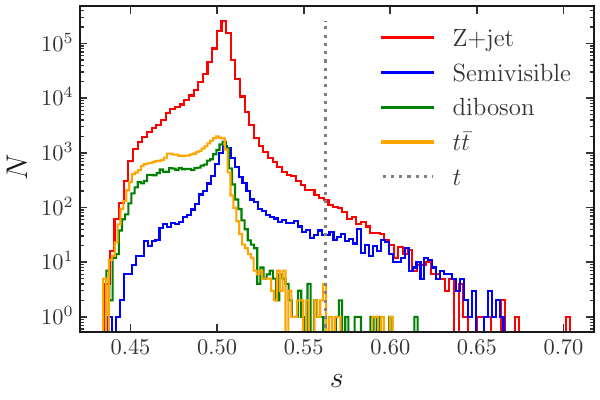}
  \includegraphics[width=0.49\textwidth]{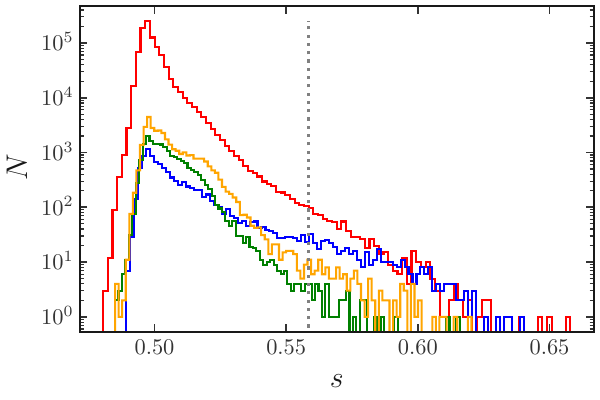}
  \caption{Same as Fig.~\ref{fig:app2} but for overestimated (left) additional backgrounds in the control region ($r^{\rm CR}_{t\bar{t}} = 5.0\%$ and $r^{\rm CR}_{VV} = 3.0\%$) and underestimated (right) additional backgrounds ($r^{\rm CR}_{t\bar{t}} = 2.8\%$ and $r^{\rm CR}_{VV} = 1.6\%$). We also show the threshold value $t$ for $\epsilon^{\rm CR}=0.001$ as a vertical line.}
  \label{fig:app3}
\end{figure}

In Fig.~\ref{fig:app2} and Fig.~\ref{fig:app3}, we show the signal score $s$ including subdominant backgrounds as discussed in Sec.~\ref{sec:Realitychecks}. The signal fraction is always fixed at $f^{\rm SR}=1\%$ (and $f^{\rm CR}=0$). We use the nominal fractions $r^{\rm SR}_{t\bar t}=3.5\%$ and $r^{\rm SR}_{VV}=2\%$ for the subdominant backgrounds in the signal region and vary $r^{\rm CR}_{t\bar t}$ and $r^{\rm CR}_{VV}$ in the control region. If the mismodelling is too severe (right plot in Fig.~\ref{fig:app2}), the classifier mainly tags $t\bar t$ and diboson events as signal-like, as expected and also shown in Tab.~\ref{tab:result5}. If the mismodelling is reduced (right plot in Fig.~\ref{fig:app3}) the semi-visible events dominate at large scores as it should be. This is even more so the case for perfect modelling (left plot in Fig.~\ref{fig:app2}). The left plot in Fig.~\ref{fig:app3} shows that slightly overestimating distinct (multi-prong) jets in the control region is less dangerous to the CWoLa method than an underestimation. In this case, these jets are more abundant in the (not perfectly modeled) control region leading to scores $s<0.5$.

\newpage

\end{appendix}

\bibliographystyle{JHEP_improved}
\bibliography{bibliography.bib}

\end{document}